\documentclass[longauth,letter,structabstract]{aa} 
\usepackage{graphicx}
\usepackage{txfonts}
\usepackage{natbib}
\def\thco{$^{13}$CO}
\def\hho{H$_2$O}
\def\hh{H$_2$}
\def\hcop{HCO$^+$}
\def\kms{km\,s$^{-1}$}
\def\pow#1#2{#1$\times$10$^{#2}$}
\def\scm{cm$^{-2}$}
\def\ccm{cm$^{-3}$}
\def\msol{M$_{\odot}$}
\def\lsol{L$_{\odot}$}
\def\dv{$\Delta$V}
\def\vlsr{$V_{\rm LSR}$}
\def\gtsim{{_>\atop{^\sim}}}
\def\ltsim{{_<\atop{^\sim}}}
\def\nhh{$n$(H$_2$)}
\def\tkin{$T_{\rm kin}$}
\def\new#1{{#1}}
\begin{document}
\title{Water abundance variations around high-mass protostars: HIFI observations of the DR21 region\thanks{\textit{Herschel} is an ESA space observatory with science instruments provided
by European-led Principal Investigator consortia and with important participation from NASA}}
\titlerunning{HIFI observations of \thco\ and \hho\ toward DR21}


\author{F.F.S. van der Tak \inst{\ref{sron},\ref{rug}} \and
        M.G. Marseille  \inst{\ref{sron}} \and
        F. Herpin \inst{\ref{bordeaux}} \and
        F. Wyrowski \inst{\ref{mpifr}} \and
        A. Baudry \inst{\ref{bordeaux}} \and
        S. Bontemps \inst{\ref{bordeaux}} \and
        J. Braine \inst{\ref{bordeaux}} \and
        S. Doty \inst{\ref{denison}} \and
        W. Frieswijk  \inst{\ref{rug}} \and
        G. Melnick \inst{\ref{cfa}} \and
        R. Shipman  \inst{\ref{sron}} \and
        E.F. van Dishoeck \inst{\ref{leiden},\ref{mpe}} \and
        A.O. Benz  \inst{\ref{eth}} \and
        P. Caselli\inst{\ref{leeds}}
        M. Hogerheijde\inst{\ref{leiden}} \and
        D. Johnstone\inst{\ref{hia},\ref{uvic}} \and
        R. Liseau\inst{\ref{onsala}} \and
        R. Bachiller\inst{\ref{oan}} \and
        M. Benedettini\inst{\ref{rome}} \and
        E. Bergin\inst{\ref{umich}} \and
        P. Bjerkeli\inst{\ref{onsala}} \and
        G. Blake\inst{\ref{caltech}} \and
        S. Bruderer\inst{\ref{eth}} \and
        J. Cernicharo\inst{\ref{csic}} \and
        C. Codella\inst{\ref{rome}} \and
        F. Daniel\inst{\ref{meudon},\ref{csic}} \and
        A.M. di Giorgio\inst{\ref{rome}} \and
        C. Dominik\inst{\ref{amsterdam}} \and
        P.~Encrenaz\inst{\ref{lerma}} \and
        M.~Fich\inst{\ref{waterloo}} \and
        A.~Fuente\inst{\ref{oan}} \and
        T.~Giannini\inst{\ref{rome}} \and
        J. Goicoechea\inst{\ref{csic}} \and
        Th. de Graauw  \inst{\ref{alma}} \and
        F. Helmich \inst{\ref{sron}} \and
        G. Herczeg\inst{\ref{mpe}} \and
        J. J{\o}rgensen\inst{\ref{kopenhagen}} \and
        L. Kristensen\inst{\ref{leiden}} \and
        B. Larsson\inst{\ref{stockholm}} \and
        D. Lis\inst{\ref{caltech}} \and
        C. McCoey\inst{\ref{waterloo}} \and
        D. Neufeld\inst{\ref{jhu}} \and
        B. Nisini\inst{\ref{rome}} \and
        M. Olberg  \inst{\ref{onsala}} \and
        B. Parise\inst{\ref{mpifr},\ref{kosma}} \and
        J. Pearson\inst{\ref{jpl}} \and
        R. Plume\inst{\ref{calgary}} \and
        C. Risacher\inst{\ref{sron}} \and
        J. Santiago\inst{\ref{oan}} \and
        P. Saraceno\inst{\ref{rome}} \and
        M. Tafalla\inst{\ref{oan}} \and
        T. van Kempen\inst{\ref{cfa}} \and
        R. Visser\inst{\ref{leiden}} \and
        S. Wampfler\inst{\ref{eth}} \and
        U. Y{\i}ld{\i}z\inst{\ref{leiden}} \and
        L. Ravera  \inst{\ref{toulouse}} \and
        P. Roelfsema  \inst{\ref{sron}} \and
        O. Siebertz \inst{\ref{kosma}} \and
        D. Teyssier  \inst{\ref{esa}}
}

\institute{SRON Netherlands Institute for Space Research, Landleven 12, 9747 AD Groningen, The Netherlands; \email{vdtak@sron.nl} \label{sron} \and
          Kapteyn Institute, University of Groningen, The Netherlands \label{rug} \and
          Laboratoire d'Astrophysique de Bordeaux, Floirac, France \label{bordeaux} \and
          Max-Planck-Institut f\"ur Radioastronomie, Bonn, Germany \label{mpifr} \and
          Sterrewacht, Universiteit Leiden, The Netherlands \label{leiden} \and
          KOSMA, I. Physik. Institut, Universit\"at zu K\"oln, Germany \label{kosma} \and
          CESR, Universit\'e de Toulouse, France \label{toulouse} \and
          Chalmers University of Technology, 41296 G\"oteborg, Sweden \label{onsala} \and
          Institute of Astronomy, ETH Z\"urich, 8093 Z\"urich, Switzerland \label{eth} \and    
          Joint ALMA Observatory, Santiago, Chile \label{alma} \and
          European Space Astronomy Centre, ESA, Madrid, Spain \label{esa} \and
          Harvard-Smithsonian Center for Astrophysics, Cambridge, USA \label{cfa} \and
Denison University, Granville OH, USA \label{denison} \and
MPI f\"ur Extraterrestrische Physik, Garching, Germany \label{mpe} \and
School of Physics and Astronomy, University of Leeds, UK \label{leeds} \and
Herzberg Institute of Astrophysics, Victoria, Canada \label{hia} \and
Dept of Physics and Astronomy, University of Victoria, Canada \label{uvic} \and
Observatorio Astron\'{o}mico Nacional, Alcal\'{a} de Henares, Spain \label{oan} \and
INAF - Istituto di Fisica dello Spazio Interplanetario, Roma, Italy \label{rome} \and
Dept of Astronomy, University of Michigan, Ann Arbor, USA \label{umich} \and
California Institute of Technology, Pasadena CA 91125, USA \label{caltech} \and
CAB, INTA-CSIC, Torrej\'{o}n de Ardoz, Spain \label{csic} \and
University of Amsterdam, The Netherlands \label{amsterdam} \and
LERMA and UMR 8112 du CNRS, Observatoire de Paris, France \label{lerma} \and
Dept of Physics and Astronomy, University of Waterloo, Canada \label{waterloo} \and
Centre for Star and Planet Formation, U. of Copenhagen, Denmark \label{kopenhagen} \and
Department of Astronomy, Stockholm University, Sweden \label{stockholm} \and
Johns Hopkins University, Baltimore, USA \label{jhu} \and
JPL, California Institute of Technology, Pasadena, CA 91109, USA \label{jpl} \and
Dept of Physics and Astronomy, University of Calgary, Canada \label{calgary} \and
Observatoire de Paris-Meudon, Meudon, France \label{meudon}}

\date{Received 26 March 2010 ; accepted 20 April 2010 }

\abstract
{Water is a key molecule in the star formation process, but its spatial distribution in star-forming regions is not well known.}
{We study the distribution of dust continuum and \hho\ and \thco\ line emission in DR21, a luminous star-forming region with a powerful outflow and a compact \ion{H}{ii} region.}
{\textit{Herschel}-HIFI spectra near 1100\,GHz show narrow \thco\ 10$-$9 emission and \hho\ $1_{11}-0_{00}$ absorption from the dense core and broad emission from the outflow in both lines. The \hho\ line also shows absorption by a foreground cloud known from ground-based observations of low-$J$ CO lines.}
{The dust continuum emission is extended over 36$''$ FWHM, while the \thco\ and \hho\ lines are confined to $\approx$24$''$ or less. The foreground absorption appears to peak further North than the other components. Radiative transfer models indicate \new{very low} abundances of $\sim$\pow{2}{-10} for \hho\ and $\sim$\pow{8}{-7} for \thco\ in the dense core, \new{and higher \hho\ abundances of $\sim$\pow{4}{-9} in the foreground cloud and $\sim$\pow{7}{-7} in the outflow}.}
{The high \hho\ abundance in the warm outflow is probably due to the evaporation of water-rich icy grain mantles, \new{while the \hho\ abundance is kept down by freeze-out in the dense core and by photodissociation in the foreground cloud}.}


\maketitle

\section{Introduction}
\label{s:intro}

The water molecule is a key species throughout the formation of stars and planets. In the gas phase, it acts as a coolant of collapsing interstellar clouds; in the solid state, it acts as glue for dust grains in protoplanetary disks to make planetesimals; and as a liquid, it acts as transporter bringing molecules together on planetary surfaces, a key step towards biogenic activity.
The first role is especially important for high-mass star formation which depends on the balance between the collapse of a massive gas cloud and its fragmentation \citep{review}.

Interstellar \hho\ is well known from ground-based observations of the 22\,GHz maser line. 
Previous space-based submm and far-IR observations have measured \hho\ abundances ranging from 10$^{-8}$ in cold gas to 10$^{-4}$ in warm gas (ISO: \citealt{iso}; SWAS: \citealt{swas}; Odin: \citealt{odin})
but did not have sufficient angular resolution to determine the spatial distribution of \hho. 
In contrast, space-based mid-IR and ground-based mm-wave observations have high angular resolution but only probe the small fraction of the gas at high temperatures (\citealt{iram}; \citealt{spitzer}).  

This paper presents observations of an \hho\ ground state line at $>$3$\times$ higher angular resolution than previously possible for such lines.
Through radiative transfer models, we compare the abundance distribution of \hho\ with that of \thco\ and dust.
%
The source DR21 (Main) is a high-mass protostellar object ($L$=45,000\,\lsol) located in the Cygnus~X region at $d$=1.7\,kpc \citep{schneider}, about 3$'$ South of the well-known DR21(OH) object (also known as W75S).
Maps of the 1.2\,mm dust emission show a dense core with a mass of 600--1000\,\msol\ and a size of 0.19$\times$0.14\,pc FWHM, surrounded by an extended envelope with mass 4750\,\msol\ and size 0.3~pc \citep{motte}.
Gas densities of 10$^5$--10$^6$\,\ccm\ are derived from both the mm-wave continuum and HCN and HCO$^+$ line emission \citep{kirby}.
Signs of active high-mass star formation are the bright mid-IR emission (272\,Jy at 21\,$\mu$m), the presence of an \hho\ 22\,GHz maser (see catalog of \citealt{braz}) and emission from ionized gas extending over 20--30$''$ \citep{pjotr}.
Together with the powerful molecular outflow \citep{garden} these signs indicate that the source is relatively evolved within the embedded phase of high-mass star formation, beyond the `ultracompact \ion{H}{ii} region' phase.

\begin{figure*}[tb]
\centering
\includegraphics[width=6cm,angle=0]{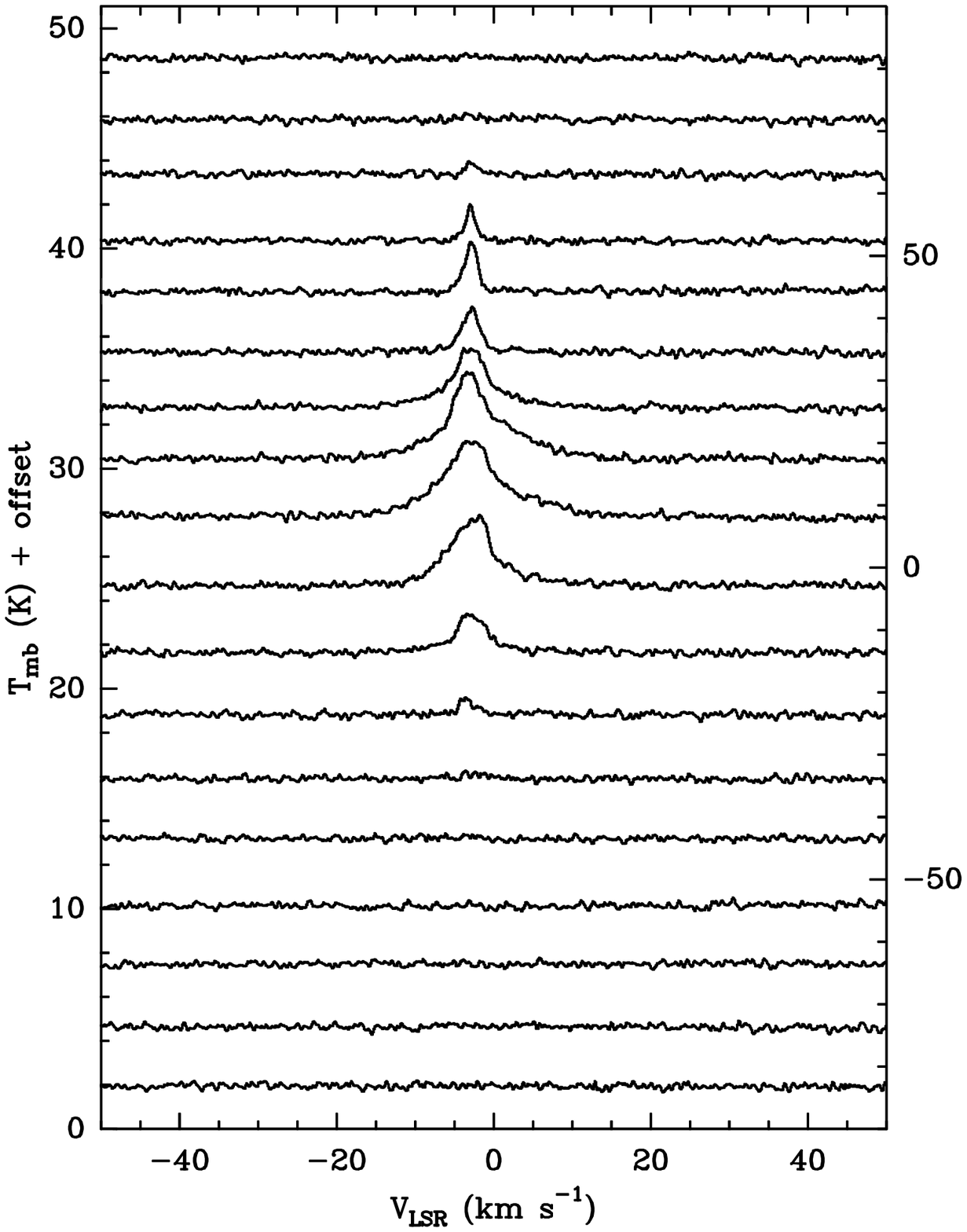}
\includegraphics[width=6cm,angle=0]{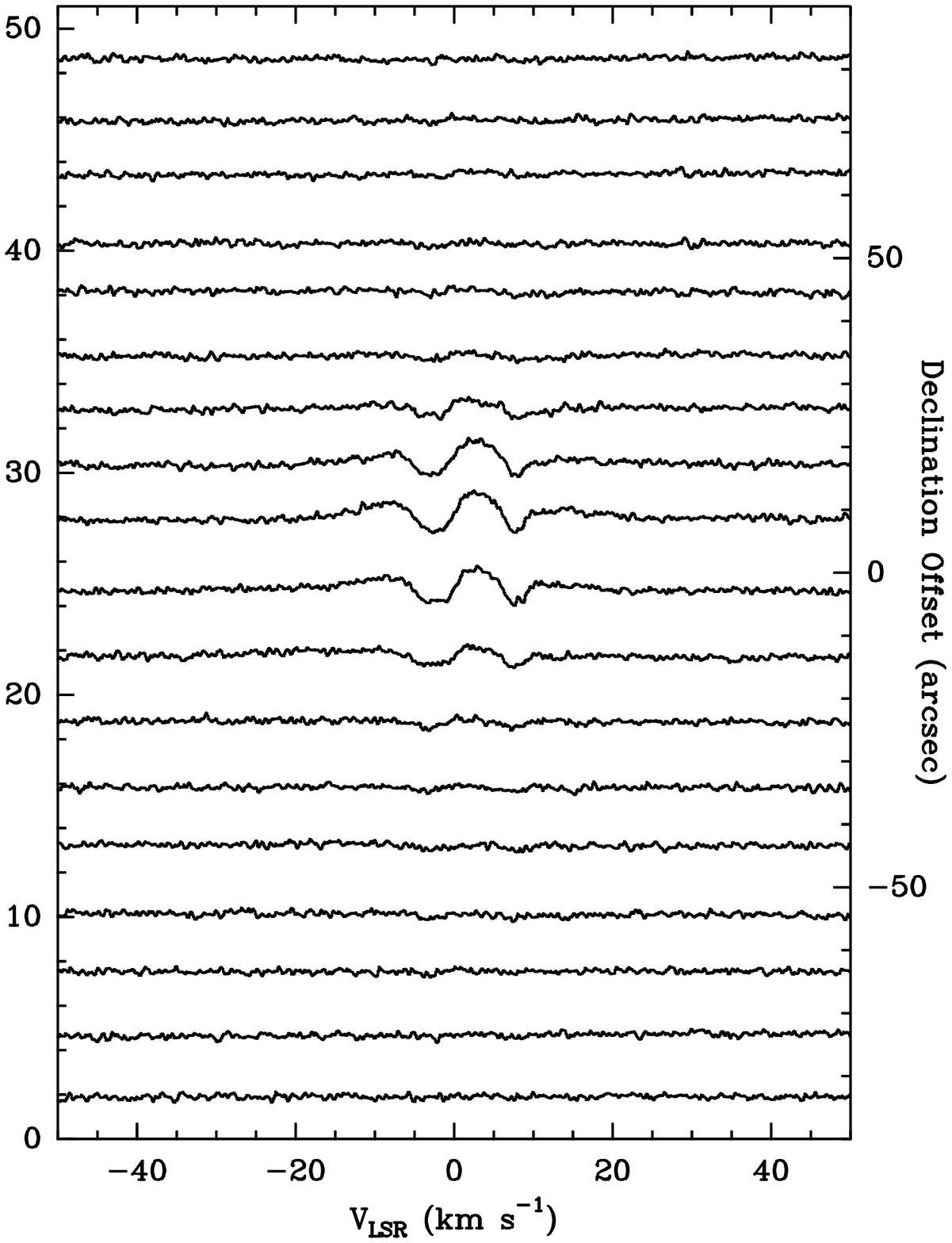}
\caption{Spectra of \thco\ $10-9$ (left) and \hho\ $1_{11}-0_{00}$ (right) lines toward DR21, taken with the WBS backend.}
\label{f:wbs}
\end{figure*} 

\section{Observations}
\label{s:obs}

The DR21 region was observed with the Heterodyne Instrument for the Far-Infrared (HIFI; De Graauw et al, this volume) onboard ESA's \textit{Herschel} Space Observatory (Pilbratt et al, this volume) on June 22, 2009.
Spectra were taken in double sideband mode using receiver band 4b, with $\nu_{\rm LO}$=1107.990\,GHz and $\nu_{\rm IF}$=6\,GHz.
The data were taken during the performance verification (PV) phase using the double beam switch observing mode with a throw of $2\farcm5$ to the SW. 
The position observed is R.A. 20:39:02.38, Dec +42:19:33.5 (J2000), close to radio peak C from \citet{pjotr}. 
A strip map was made in the N-S direction, spanning offsets from +90$''$ to $-$90$''$ at a $10\farcs5$ spacing, half the beam size of 21$''$ FWHM at our observing frequency, which corresponds to 0.17\,pc at the distance of DR21. This beam size was measured before launch and is 10\% larger than the diffraction limit due to spillover effects.

Data were taken with two backends: the acousto-optical Wide-Band Spectrometer (WBS) which covers 1140\,MHz bandwidth at 1.1\,MHz (0.30\,\kms) resolution, and the correlator-based High-Resolution Spectrometer (HRS), which covers 230\,MHz bandwidth at 0.48\,MHz (0.13\,\kms) resolution. Two polarizations are available except for the HRS data of \hho.

The system temperature of our data is 340--360\,K \new{DSB} and the integration time is 67\,seconds per position (ON+OFF). 
Calibration of the raw data onto $T_A^*$ scale was performed by the in-orbit system (Roelfsema et al, in prep); conversion to $T_{mb}$ was done assuming a beam efficiency of 0.67 as estimated by the Ruze formula and validated by raster maps of Saturn (M. Olberg, priv. comm.). Currently, the flux scale is accurate to $\approx$10\% which will improve when the telescope efficiency and sideband ratio are measured on Mars.
The calibration of the data was performed in the \textit{Herschel} Interactive Processing
Environment (HIPE; \citealt{ott}) version 2.1; further analysis was done within the CLASS\footnote{\tt
  http://www.iram.fr/IRAMFR/GILDAS} package. 
After inspection, data from the two polarizations were averaged together to obtain rms noise levels of 97\,mK on 0.5\,MHz channels for the WBS data, 195\,mK on 0.24\,MHz channels for the \thco\ HRS data and 244\,mK on 0.24\,MHz channels for the \hho\ HRS data.

\section{Results}
\label{s:res}

Figure~\ref{f:wbs} shows the WBS spectra from both receiver sidebands at each of the 18 offset positions.
The spectra from both backends show a continuum signal and two line features. The first line feature is seen in emission at $\nu$ = 1101.357\,GHz LSB or 1114.623\,GHz USB, which we identify with the \thco\ $J$=10$\to$9 line at 1101.349597\,GHz\footnote{Spectroscopic data are taken from the CDMS catalog \citep{cdms} at {\tt http://cdms.de}}. This line has an upper level energy ($E_u$) of 291\,K and a critical density of \pow{1}{6}\,\ccm, using collision data from \citet{flower}.
The second feature has a mixed emission-absorption profile and lies at $\nu$ = 1102.423\,GHz LSB or 1113.557\,GHz USB, which we identify with the \hho\ $J_{\rm K_pK_o}$=$1_{11}\to0_{00}$ line at 1113.34306\,GHz. This line has $E_u$ = 53\,K and a critical density of \pow{3-6}{8}\,\ccm, using collision data from \citet{faure}. 
%
%
No other lines are detected.

\begin{figure}[tb]
\centering
\includegraphics[width=4cm,angle=-90]{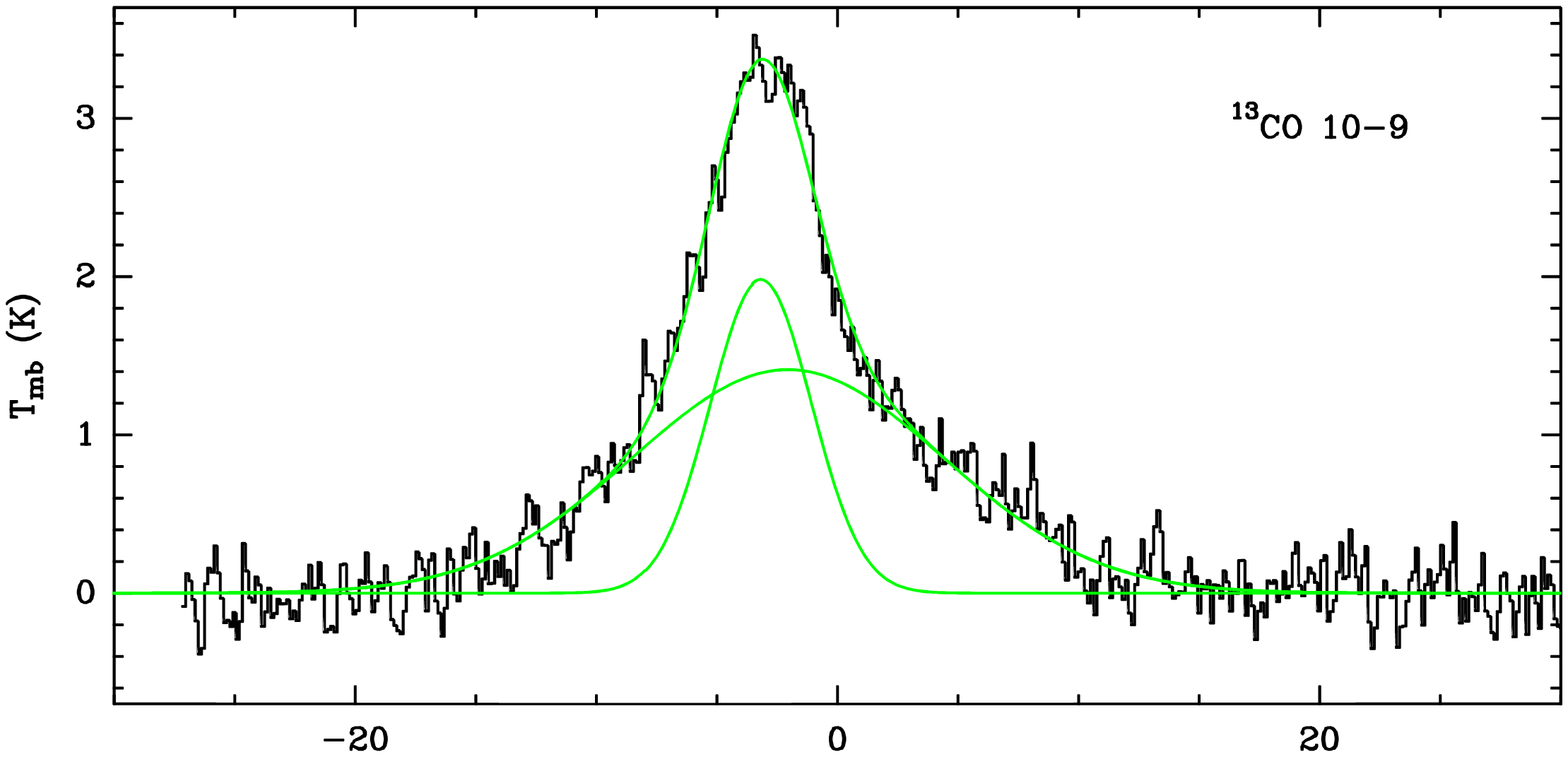}  
\includegraphics[width=4.5cm,angle=-90]{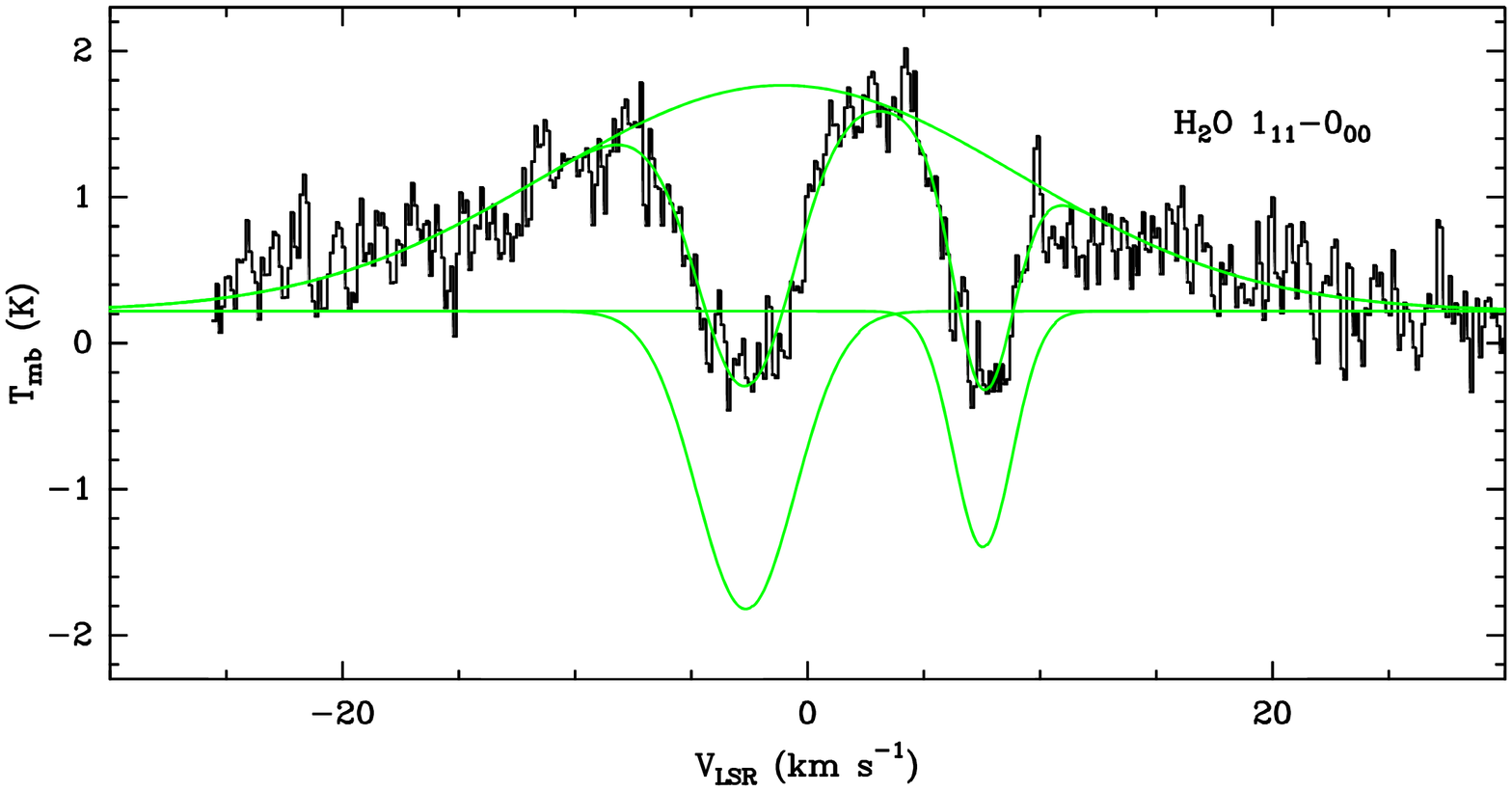}  
\caption{Spectra of \thco\ $10-9$ (top) and \hho\ $1_{11}-0_{00}$ (bottom) lines toward the central position, taken with the HRS backend, with Gaussian decompositions overplotted.}
\label{f:hrs}
\end{figure} 

Figure~\ref{f:hrs} shows the HRS spectra of the \thco\ and \hho\ lines at the central position, with the continuum subtracted. 
The \thco\ line profile has a double structure which is seen to be well reproduced by the sum of two Gaussians: a narrow component (\dv$=$4.9\,\kms) centered at \vlsr$=$--3.2\,\kms, which we attribute to the molecular cloud core (protostellar envelope), and a broad component (\dv$=$15.2\,\kms) centered at \vlsr$=$--2.0\,\kms, which we ascribe to the molecular outflow.
In contrast, the \hho\ line profile shows two absorption features, with strong emission in between, and weaker emission on the far blue- and redshifted sides. Overplotted is a decomposition with three Gaussian components: broad (\dv$=$23.8\,\kms) emission centered at \vlsr=--1.0\,\kms, and absorptions at \vlsr=--2.6\,\kms\ and +7.6\,\kms\ with \dv$=$5.0 and 3.0\,\kms.
This shape resembles the prediction by \citet{poelman} for the \hho\ ground state lines towards high-mass protostellar envelopes for the case of a constant, low \hho\ abundance (\hho/\hh\ $\sim$10$^{-9}$) with two modifications. First is the broad emission also seen in \thco\ which is likely due to the outflow. Second is the absorption at \vlsr=7.6\,\kms\ which is known from ground-based observations of low-$J$ CO lines \citep{jakob} and likely due to a foreground cloud. This absorption is not seen in the \thco\ 10-9 line, nor in ground-based observations of mid-$J$ HCN and \hcop\ lines, which indicates a low temperature ($\sim$10\,K) and density ($\ltsim$10$^4$\,\ccm) for the foreground cloud. Both absorptions are also seen in the SWAS spectra of the o-\hho\ and C$^0$ ground-state lines, but not in \thco\ 5--4 \citep{ashby}.

\begin{figure}[tb]
\centering
\includegraphics[width=4cm,angle=0]{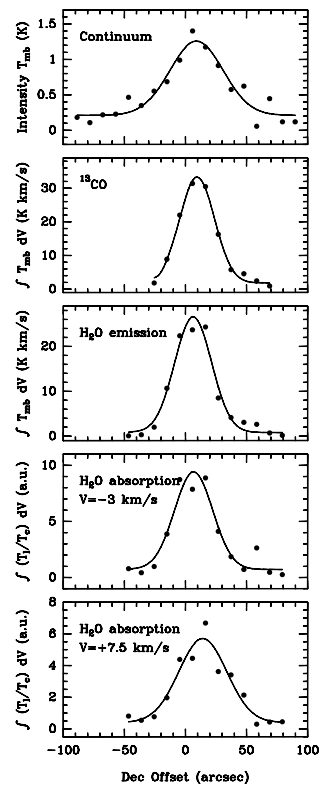}
\caption{Plots of observed intensity versus spatial offset with Gaussian models superposed. }
\label{f:spatial}
\end{figure}

Figure~\ref{f:spatial} shows the spatial brightness distribution of the continuum and the lines. The points are the results of Gaussian profile fits for the lines, and of linear baseline fits for the continuum, at each offset position. For the \hho\ line, separate fits were made for the emission and the two absorption components. The curves in Figure~\ref{f:spatial} are Gaussian fits to the observed spatial distribution.
The dust emission peaks at an offset of (8.7$\pm$0.5)$''$ North of the nominal position, and the emission distribution has an FWHM width of (35.7$\pm$0.5)$''$, consistent with ground-based measurements at longer wavelengths \citep{gibb,motte}. The peak of the \thco\ emission is at offset (9.65$\pm$0.5)$''$, consistent with the dust peak within the combined error, but the FWHM of the \thco\ emission is (24.1$\pm$0.4)$''$, essentially unresolved. The \hho\ emission component has an FWHM of (25.2$\pm$2.5)$''$, similar to \thco, but its peak is shifted further South, at offset (6.4$\pm$1.1)$''$.
The spatial distribution of the \hho\ absorption at $V$ = --3\,\kms\ 
is the same as that of the \hho\ emission, peaking at offset (6.7$\pm$1.6)$''$ with an FWHM of (26.2$\pm$3.3)$''$. In contrast, the \hho\ absorption at $V$ = +7.5\,\kms\ peaks much further North, at offset (14.2$\pm$1.9)$''$, and is \new{possibly} extended with an FWHM of (32.8$\pm$4.7)$''$, which is a lower limit to its full width because of insufficient background signal. Indeed, \citet{schneider} find that the W75N cloud extends over several arcminutes.
%

\section{Discussion and Conclusions}
\label{s:disc}


To estimate the \hho\ and \thco\ abundances from our data, we have run spherical radiative transfer models following \citet{marseille}.
First, the dust continuum emission was modeled with the MC3D program \citep{mc3d}
with the source size and luminosity kept fixed at the values in \S\ref{s:intro}.
The continuum data are consistent with a power-law density profile $n=n_0 (r/r_0)^{-\alpha}$ with the index $\alpha$=1.5 as expected for evolved protostellar envelopes \citep{massive}. 
Derived temperatures range from 117\,K at the adopted inner radius of 0.01~pc to 23\,K at the outer radius of 0.3~pc; densities drop from \pow{3}{7}\,\ccm\ to \pow{2}{5}\,\ccm.
This temperature and density profile was adopted for the line radiative transfer with the RATRAN program \citep{ratran}.
The abundance of \hho\ was varied between 10$^{-10}$ and 10$^{-7}$ and the \thco\ abundance between 10$^{-7}$ and 10$^{-6}$, both independent of radius.

\begin{figure}[tb]
\centering
\includegraphics[width=3.5cm,angle=-90]{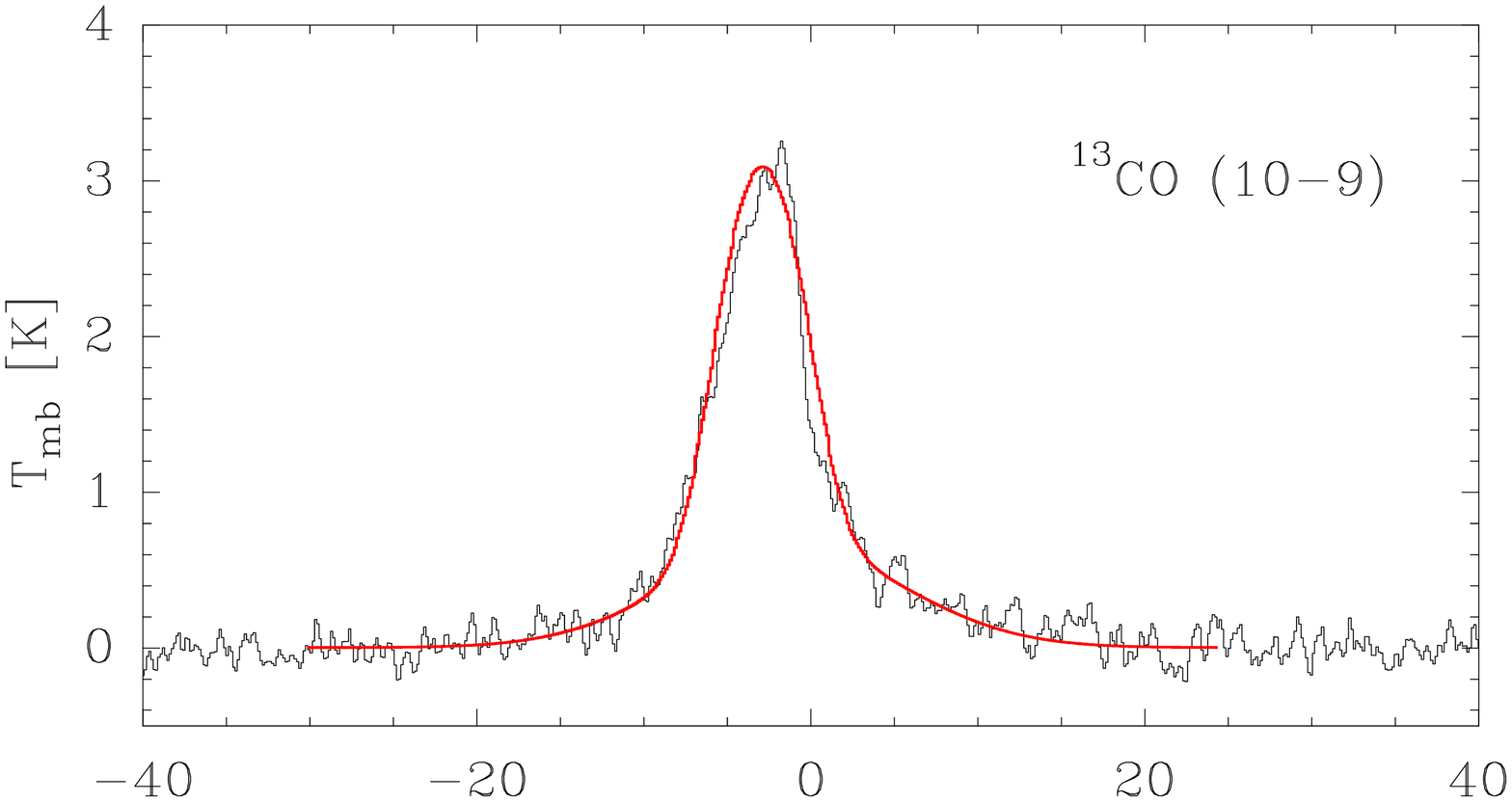}
\includegraphics[width=4.5cm,angle=-90]{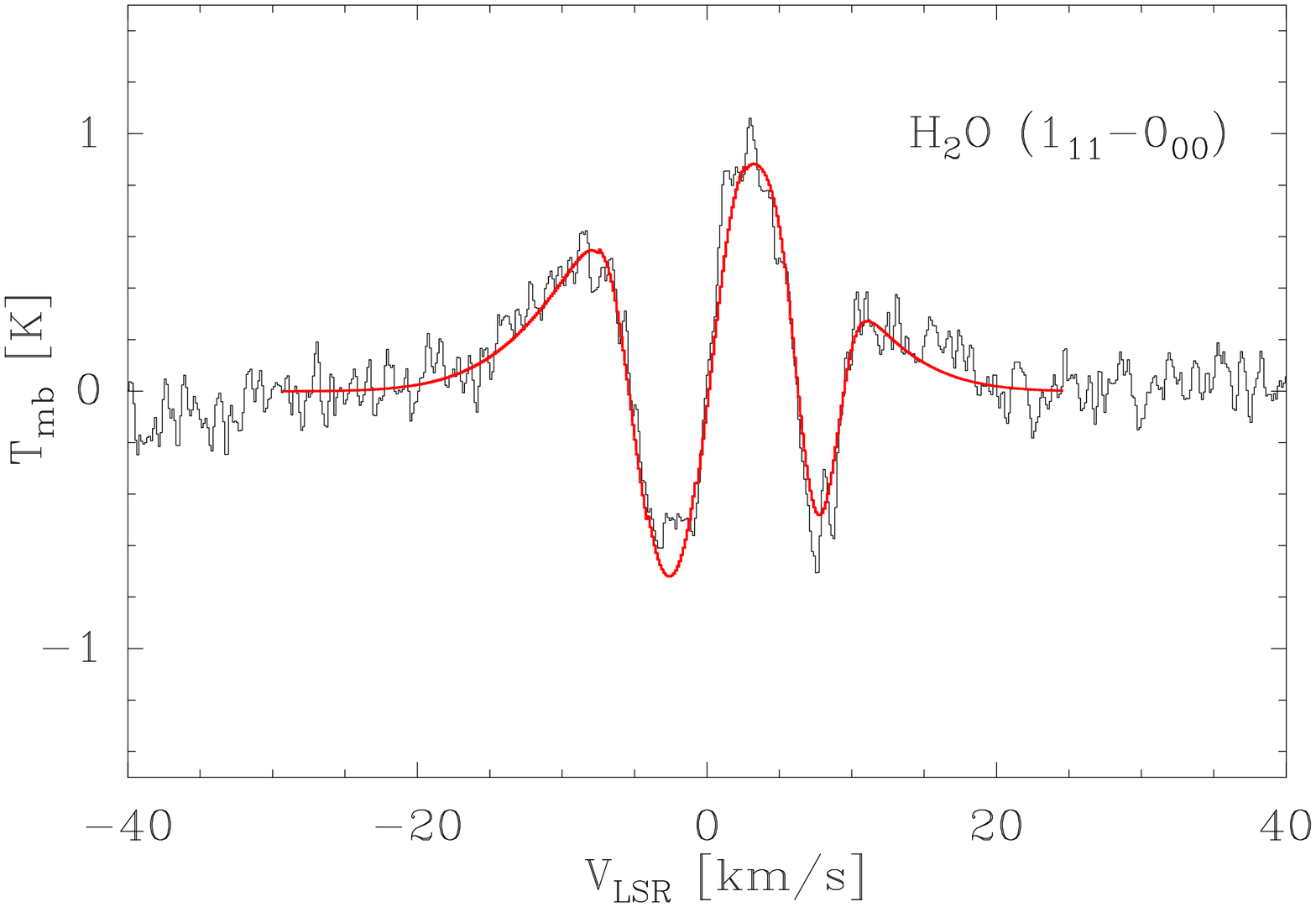}
\caption{Data from Fig.~\ref{f:hrs} with radiative transfer models superposed. }
\label{f:ratran}
\end{figure}

The red lines in Figure~\ref{f:ratran} show the results of our best-fit model. The fit to the line profiles at the central few positions is good if two components are added to the dense core model: one representing the outflow and one for the foreground cloud. 
This simple addition procedure is valid because the foreground cloud is transparent and the outflow is seen at an angle, so that neither component blocks the view of the dense core.
The modeled optical depths are 2.8 for the \hho\ line and 0.77 for \thco, and the derived abundances are uncertain to $\approx$35\% which is the quadratic sum of 30\% from the line intensity and 20\% from the core mass estimate. 
The model underproduces the observed line strengths at large position offsets, which suggests that the density profile flattens out at large radii, as also indicated by the observed continuum brightness at the offset positions.

Deriving the \thco\ and \hho\ abundances in the outflow and the foreground using radiative transfer models is not possible because the masses and \hh\ column densities of these components are unknown. Instead we have used RADEX \citep{radex} to estimate their \thco\ and \hho\ column densities, \new{and use a $^{12}$C/$^{13}$C ratio of 60 \citep{isotope} and a CO abundance of \pow{2}{-4} \citep{lacy} to convert $N$(\thco) to $N$(\hh) and estimate $x$(\hho) as $N$(\hho)/$N$(\hh)}. For the foreground cloud, we adopt \tkin = 10\,K and \nhh = 10$^4$\,\ccm, and for the outflow, we adopt \tkin = 200\,K and \nhh = \pow{3}{4}\,\ccm, \new{which assumptions introduce a factor of $\sim$2 uncertainty each}. To estimate $N$(\thco) for the foreground cloud, we use the $J$=1-0 observations by \citet{jakob}.

Table~\ref{t:abs} summarizes our derived column densities and abundances of \thco\ and \hho\ in the various physical components of the DR21 region. 
\new{Our \hho\ abundance in the dense core is $\sim$100$\times$ lower than previous determinations (\S\ref{s:intro}) but should be regarded as a lower limit.}
At the low temperatures and high densities in the core, most \hho\ is likely frozen on grains, and the observed line may arise in a small region with a high \hho\ abundance. 
The derived \thco\ abundance for the core is $\sim$4$\times$ lower than expected for the above values of the CO isotopic ratios and abundance, which suggests that even some CO is frozen out in the outer parts of the core.
The density of the foreground cloud is too low for significant freeze-out, but with $A_V$ $\approx$1.2\,mag, photodissociation is rapid for \hho\ but not for \thco.

The high \hho\ abundance for the outflow is likely related to its
temperature of $\sim$200\,K, which is high enough to have \hho\ released from the dust grains by thermal evaporation, or possibly by shocks \citep{melnick}. Further enhancement may be expected in even warmer gas ($\gtsim$250\,K) when neutral-neutral reactions drive most gas-phase oxygen into \hho, but such gas is not probed by our data. Future HIFI observations of high-excitation \hho\ lines towards protostars of all masses will however very likely reveal this effect.

\begin{table}
\caption{Abundances and column densities of \thco\ and \hho.}
\label{t:abs}
\begin{tabular}{ccccc}
\hline \hline
\noalign{\smallskip}
Component & $N$(\thco)$^a$  & $N$(p-\hho)     & $N$(\hh) & $x$(p-\hho)\\
          & 10$^{16}$\,\scm & 10$^{12}$\,\scm & 10$^{21}$\,\scm & \\ 
\noalign{\smallskip}
\hline
\noalign{\smallskip}
Dense core & \pow{7.8}{-7} & ...              & ... & \pow{1.6}{-10} \\
Outflow    & 5             & 10,000           & 15  & \pow{7}{-7} \\
Foreground & 0.7           & 4                & 2.1 & \pow{4}{-9} \\
\noalign{\smallskip}
\hline
\noalign{\smallskip}
\multicolumn{5}{l}{$^a$: Abundance for the dense core} \\
\end{tabular}
\end{table}

\begin{acknowledgements}
HIFI has been designed and built by a consortium of institutes and university departments from across Europe, Canada and the US under the leadership of SRON Netherlands Institute for Space Research, Groningen, The Netherlands with major contributions from Germany, France and the US.
Consortium members are: Canada: CSA, U.Waterloo; France: CESR, LAB, LERMA, IRAM; Germany:
KOSMA, MPIfR, MPS; Ireland, NUI Maynooth; Italy: ASI, IFSI-INAF, Arcetri-INAF; Netherlands: SRON, TUD; Poland: CAMK, CBK; Spain: Observatorio Astron\'omico Nacional (IGN), Centro de Astrobiolog\'{\i}a (CSIC-INTA); Sweden: Chalmers University of Technology - MC2, RSS \& GARD, Onsala Space Observatory, Swedish National Space Board, Stockholm University - Stockholm Observatory; Switzerland: ETH Z\"urich, FHNW; USA: Caltech, JPL, NHSC.
\end{acknowledgements}

\bibliographystyle{aa}
\bibliography{dr21}

\end{document}